\documentclass[twocolumn,showkeys,floatfix]{revtex4-1}
\usepackage{natbib}
\usepackage{graphicx}
\usepackage{subfigure}

\usepackage[latin1]{inputenc}
\usepackage{amsmath}
\usepackage{setspace}
\usepackage{amssymb}
\usepackage{color}

\newcommand{\eqb}{\begin{equation}}
\newcommand{\eqe}{\end{equation}}
\newcommand{\enb}{\begin{eqnarray}}
\newcommand{\ene}{\end{eqnarray}}

\begin{document}
\title{An Algebraic Solution for the Kermack-McKendrick Model}

\author{Alexsandro M. Carvalho} \email{alexsandromc@unisinos.br}
\affiliation{Escola de Gestão e Negócios, Programa de Pós-Graduação em
  Economia, Universidade do Vale do Rio dos Sinos, Caixa Postal 275,
  93022-000 São Leopoldo RS, Brazil}

\author{Sebastián Gonçalves} \email{sgonc@if.ufrgs.br}
\affiliation{Instituto de Física, Universidade Federal do Rio Grande
  do Sul, Caixa Postal 15051, 90501-970 Porto Alegre RS, Brazil}

\begin{abstract}
We present an algebraic solution for the Susceptible-Infective-Removed
(SIR) model originally presented by Kermack-McKendrick in 1927.
Starting from the differential equation for the removed subjects
presented by them in the original paper, we re-write it in a slightly
different form in order to derive a formal solution, unless one
integration.  Then, using algebraic techniques and some well justified
numerical assumptions we obtain an analytic solution for the
integral. We compare the numerical solution of the
differential equations of the SIR model with the analytic solution
here proposed, showing an excellent agreement. Finally, the present scheme
allow us to represent analytically two key quantities: time of the
infection peak and fraction of immunized to stop the epidemic.
\end{abstract}

%epidemics dynamics, social networks, numerical simulations
%\PACS 87.23.-n \sep 87.19.Xx

\maketitle

\section{Introduction}
The Kermack-McKendrick model~\cite{Kermack1927} or commonly called the
SIR model is a cornerstone of the theoretical mathematical models
applied to the dynamic of disease spreading, or simply epidemic
models.  Presented in 1927 in the Proceeding of Royal Society A, soon
became evident that it represents an excellent frame for the
understanding of the behavior of epidemics.

The basic reproductive number $R_0 = \beta \tau$, the infection curve
(that is the asymptotic number of removed as function of $R_0$) are
concepts that found in the SIR model the basement that provide the
perfect sense, origin, and explanation.  It is true, however, that
real epidemics are hardly described strictly in terms of the SIR model
(or one of the variants like SIS, SIRS, etc). Nevertheless, it is
almost impossible to ``speak'' the language of the mathematical
description of epidemics at any level of complexity, without having
the SIR model in mind. If a physics metaphor could be applied, we
would say that the SIR model is for epidemics dynamics as the harmonic
oscillator is for physics.  But the metaphor can not be sustained at
all levels because different from the given physical example, the SIR
equations lack an analytic solution.  To obtain the time evolution of
the S, I, and R quantities we have to resort on numerical integration
of the finite differences representation of the SIR
equations. Fortunately, from the 1980, the increasing power of
affordable computers made the finding of numerical solutions a trivial
task, avoiding all the pain of obtaining them by hand or with the help
of pocket calculators.  But the present facilities do not diminish the
beauty and intellectual satisfaction of an analytic solution.  In this
contribution we present a close solution for the classical SIR model
as was formulated by Kermack-McKendrick.

Except for the SIS model, the simplest of all epidemic models, whose
differential equations, because of the $S+I = 1$ constrain, end up to
be a single and solvable one, the Riccati equation, all the others,
remarkably the SIR one, which was the originally presented by
Kermack-McKendrick, lack a close analytic solution for their
differential equations~\footnote{In PRE 2001, Newman has presented
  a mean field exact solution of the SIR model for tree like random
  networks, but no solution for the temporal evolution of the S, I
  populations}.

While analytic tools like stability analysis,
asymptotic analysis, or phase diagram analysis can be used to get some
insight into the behavior of these models, by close analytic solution
we mean the temporal evolution of the number of infectives or
susceptibles, and this is what we want to pursue in this contribution.
In doing this task we first present a solution which is exact and
closed except for one integral. To cope with the integral, regarding
the denominator as a Taylor series or polynomial, it is then represented
as a partial fraction expansion.  From that expansion, after making an
approximation which is then numerically well justified, we arrive at
the final solution of the SIR model.

\section{SIR model and the integral form}
The differential equation of the Kermack-McKendrick model are:
\begin{eqnarray}
\frac{dS}{d\tau} &=& -R_{0} SI \nonumber\\ \frac{dI}{d\tau} &=& R_{0}
SI - I \label{eq:sir} \\ \frac{dR}{d\tau} &=& I, \nonumber
\end{eqnarray}
where $S$, $I$, and $R$ represent respectively the fraction of
susceptible, infected, and removed subjects inside a population of
fixed size. This is the no vital dynamics version of the model where
the condition $S+I+R=1$ holds. $R_{0}=\beta/\gamma$ is the basic
reproduction number and $\tau=\gamma t$.  To obtain the time evolution
of the state of the epidemic it is necessary to integrate the
equations~(\ref{eq:sir}), which can not be performed in this case.
Kermack and McKendrick, from the first and third differential
equations (Eq.~\ref{eq:sir}), arrive to the following expression for
$S(R)$:
\begin{equation}
S(\tau)=S(0)e^{-R_0R(\tau)}
\label{eq:stau}
\end{equation}
in which it is assumed that at the beginning of the infection
there is an initial fraction of infected but no removed subjects,
{\em i.e.} $I(0)=i_0$, $R(0)=0$, so $S(0)=1-i_0$.
From that expression they obtain an
independent differential equation for the removed subjects:
\begin{equation}
\frac{dR}{d\tau} = I = 1 - R - S(0)e^{-R_0 R}
\end{equation}
At this point, in order to go a step further, they use a strong
approximation valid for $R \ll 1$ (only valid for initial times or
very weak epidemics). Here, we continue without approximations by now,
making a change of variables to $w = R_0(1-R)$, in terms of which, the
last equation transforms into:
\begin{equation}
\frac{dw}{d\tau} = R_0 S(0) e^{w-R_0} - w\;.
\end{equation}
Thus, we can express the three quantities $(S,I,R)$ in terms of $w$,
i.e.,
\begin{eqnarray}
S(\tau) &=& S(0)e^{w-R_{0}} \nonumber \\
I(\tau) &=& \dfrac{w}{R_{0}}-S(0)e^{w-R_{0}} \label{eq:lw} \\
R(\tau) &=& 1-\dfrac{w}{R_{0}} \nonumber
\end{eqnarray}
Therefore, the problem of solving the original $S,I,R$ differential
equations of the Kermack Mc-Kendrick model, Eq.~(\ref{eq:sir}), was
translated into solving the equation for $w$ whose formal solution can
be expressed in term of an integral:
\begin{equation}
\int\dfrac{dw}{R_{0}S(0)e^{w-R_{0}}-w}=\tau+k,
\label{eq:integral}
\end{equation}
where $k$ is a constant of integration. We will see in the next
section how we can compute this integral.

\section{Algebraic Solution}
In the previous section we have translated the SIR problem of solving
the set of differential equations into the quest of a primitive for
the integral~(\ref{eq:integral}).  However, as far as we know, there
is no such a primitive; instead we use a functional form that
represent the integrand, in the hope that the functional form can be
integrated somehow.

\begin{figure}[!t]
\centerline{\subfigure{\includegraphics[width=8.5cm]{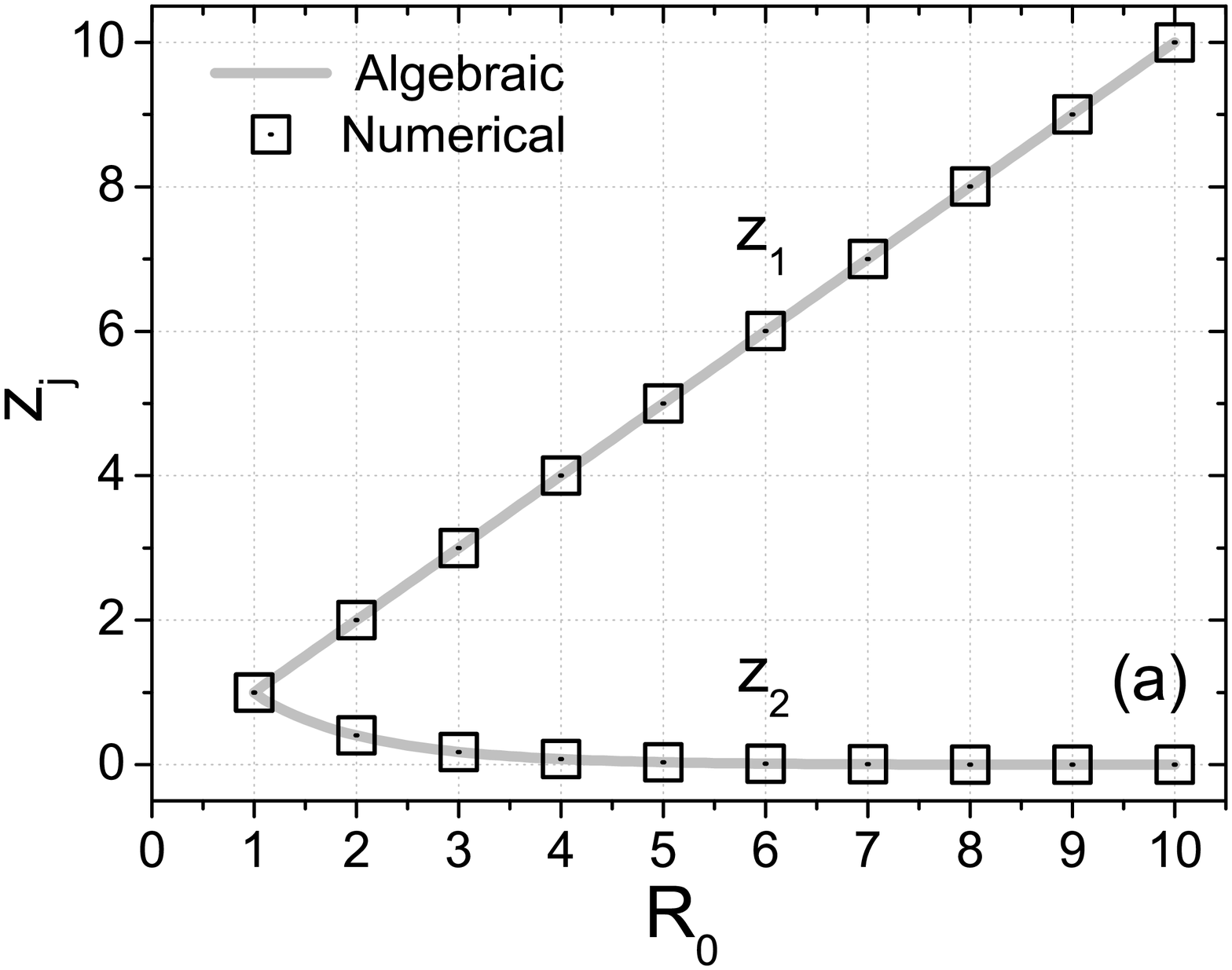}}}
\centerline{\subfigure{\includegraphics[width=7.5cm]{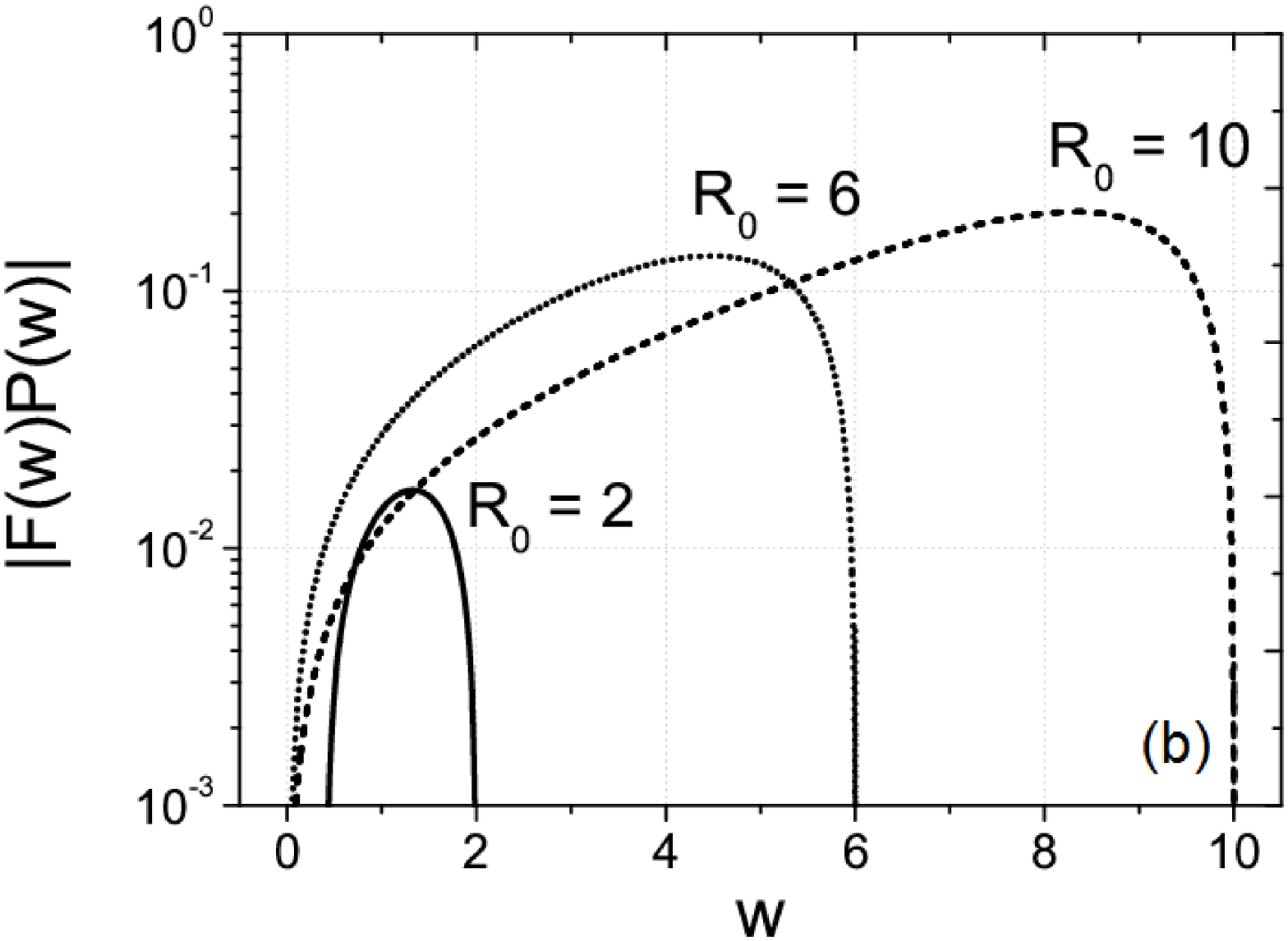}}}
\caption{(a) The two real roots of $P(w)$, $z_1$, $z_2$, as functions
  of $R_0$. Symbols represent numerical solutions of Eq.~(\ref{eq:pw})
  while the lines are the analytic relations obtained from
  Eq.~(\ref{eq:zs}). (b) Module ratio of the complex residual part,
  $F(w)$, and the integrand of Eq.~(\ref{eq:integral}), $1/P(w)$. Initial
  condition $i_0=10^{-3}$.}
\label{fig:raiz}
\end{figure}

We denote the denominator of the integrand of Eq.~(\ref{eq:integral})
as $P(w)$,
\begin{equation}
 P(w) = R_{0} S(0)e^{w-R_{0}}-w
\label{eq:pw}
\end{equation}
which can be considered, if the exponential is represented by its
Taylor expansion, as a polynomial in complex domain. Thus, its inverse
can be expanded as a partial fraction:
\begin{equation}
\frac{1}{P(w)}=\sum_{j} \dfrac{A(z_{j})}{w-z_{j}} + F(w) \;,
\label{eq:fp}
\end{equation}
where $z_j$ are the real roots of $P(w)$. The coefficients $A(z_j)$
are given by
\begin{equation}
A(z_{j})=\left(\dfrac{dP(w)}{dw}\right)^{-1}_{w=z_{j}}
=\dfrac{1}{R_{0}S(0)e^{z_{j}-R_{0}}-1}
\label{eq:Az}
\end{equation}
and $F(w)$ represents the residual part, i.e, the partial
fraction decomposition which includes the complex roots.

\begin{figure}[!t]
\centerline{\subfigure{\includegraphics[width=8.5cm]{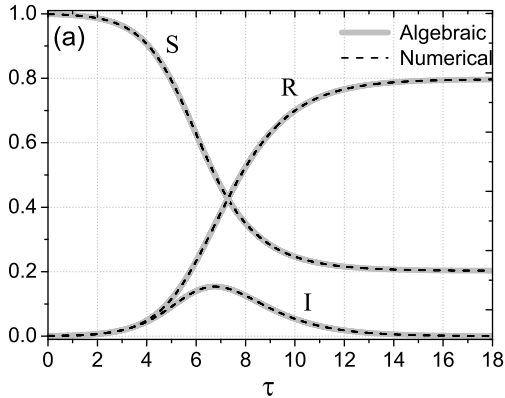}}}
\centerline{\subfigure{\includegraphics[width=8.5cm]{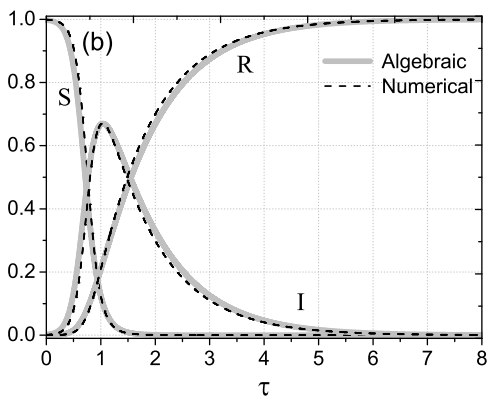}}}
\caption{Dynamics of the Susceptible-Infected-Removed (SIR) model for
  (a), $R_0=2$, and (b), $R_0=10$. Initial condition
  $i_0=10^{-3}$. Numerical solution (dashed line) refers to the
  numerical integration of the Eq.~(\ref{eq:sir}) and algebraic
  solution (solid line) refers to numerical computation of
  Eq.~(\ref{eq:lw}) and~(\ref{eq:tauf}).}
\label{fig:sir}
\end{figure}
It is straightforward to see that $P(w)$ has two real roots if $R_0 \neq 1$
(if $R_0=1$ there is one real root). Here we are only interested in
the $R_0 > 1$ cases, when the roots are related with the extreme
values of $w(\tau)$, {\em i.e} $w(0)$ and $w(\infty)$, This can be
seen using the definition of $w$ and $P(w)$, and the expression of the
susceptibles in terms of $w$ (Eq.~\ref{eq:lw}):
\begin{eqnarray}
  P(z) = R_{0} S(0)e^{z-R_{0}}-z = 0\nonumber\\
  \Rightarrow R_0 S(z) = R_0 (1 - R(z))
\label{eq:root}
\end{eqnarray}
Equivalent to have $i(\tau) = 0$, which happens at the
extreme conditions $\tau=0$ (approximately because $i(0)$ is small
but not zero) and $\tau \rightarrow \infty$ (in this case exactly).
An still approximate but much better estimation for $z_1$ can be found if we
note that for $t \ll 1$, $w(t) \approx R_0$, so $e^{w-R_{0}} \approx
1-R_0+w$ is a very good approximation, which let us arrive to the
following two real roots of $P(w)$ that we were looking for:
\begin{eqnarray}
z_1 &\approx& \frac {R_0 S(0) \left(R_0-1\right)}{R_0 S(0)-1} \nonumber \\
z_2 &=& w(\infty) = R_0 (1-R(\infty)) = R_0 S(\infty)\; ,
\label{eq:zs}
\end{eqnarray}
both of them in terms of $S(0)$ and $R_0$, and $S(\infty)$ which in
turn can be implicitly expressed using Eq.~\ref{eq:stau} in terms of
the first two quantities
\begin{equation}
S(\infty)=S(0) e^{-R_0 (1 - S(\infty))}
\label{eq:sinf}
\end{equation}

Therefore, we can write the Eq.~(\ref{eq:fp}) as
\begin{equation}
\dfrac{1}{P(w)} \approx \frac{A(z_1)}{w-z_1}+\frac{A(z_2)}{w-z_2}+F(w)
\label{eq:apw}
\end{equation}

In order to verify the contribution of $F(w)$, we evaluate numerically
the module ratio between $F(w)$ and $1/P(w)$,
$\left|F(w)P(w)\right|$. It can be seen in Fig.~\ref{fig:raiz}(b) that
that ratio is small, then from this point we neglect $F(w)$, arriving
to the following expression for $w$:
\begin{eqnarray}
\tau + k &=& \int\frac{dw}{P(w)} \nonumber\\ &\simeq&
A(z_1)\ln|w-z_1|+A(z_2)\ln|w-z_2|
\end{eqnarray}
We can give explicit expression of the coefficients $A(z_1)$ and
$A(z_2)$ from Eq.~\ref{eq:Az} and Eq.~\ref{eq:zs}:
\begin{eqnarray}
  A(z_1) &=& \dfrac{1}{R_0 S(0) e^{\frac{R_0(1-S(0))}{R_0S(0)-1}} - 1}
             \equiv A_1 \nonumber \\
  A(z_2) &=& \dfrac{1}{R_0 S(0) e^{-R_0(1-S(\infty))} - 1}
             \equiv A_2
\end{eqnarray}
And the constant $k$ is obtained form the initial condition $w(0)=R_0$.
So finally, we can approximately
determine the integral (\ref{eq:integral}) as
\begin{equation}
  \tau \simeq  A_1 \ln{\left|\frac{w-z_1}{R_0 - z_1}\right|} +
               A_2 \ln{\left|\frac{w-z_2}{R_0 - z_2}\right|}
\label{eq:tauf}
\end{equation}

\begin{figure}[!t]
\centerline{\subfigure{\includegraphics[width=8.5cm]{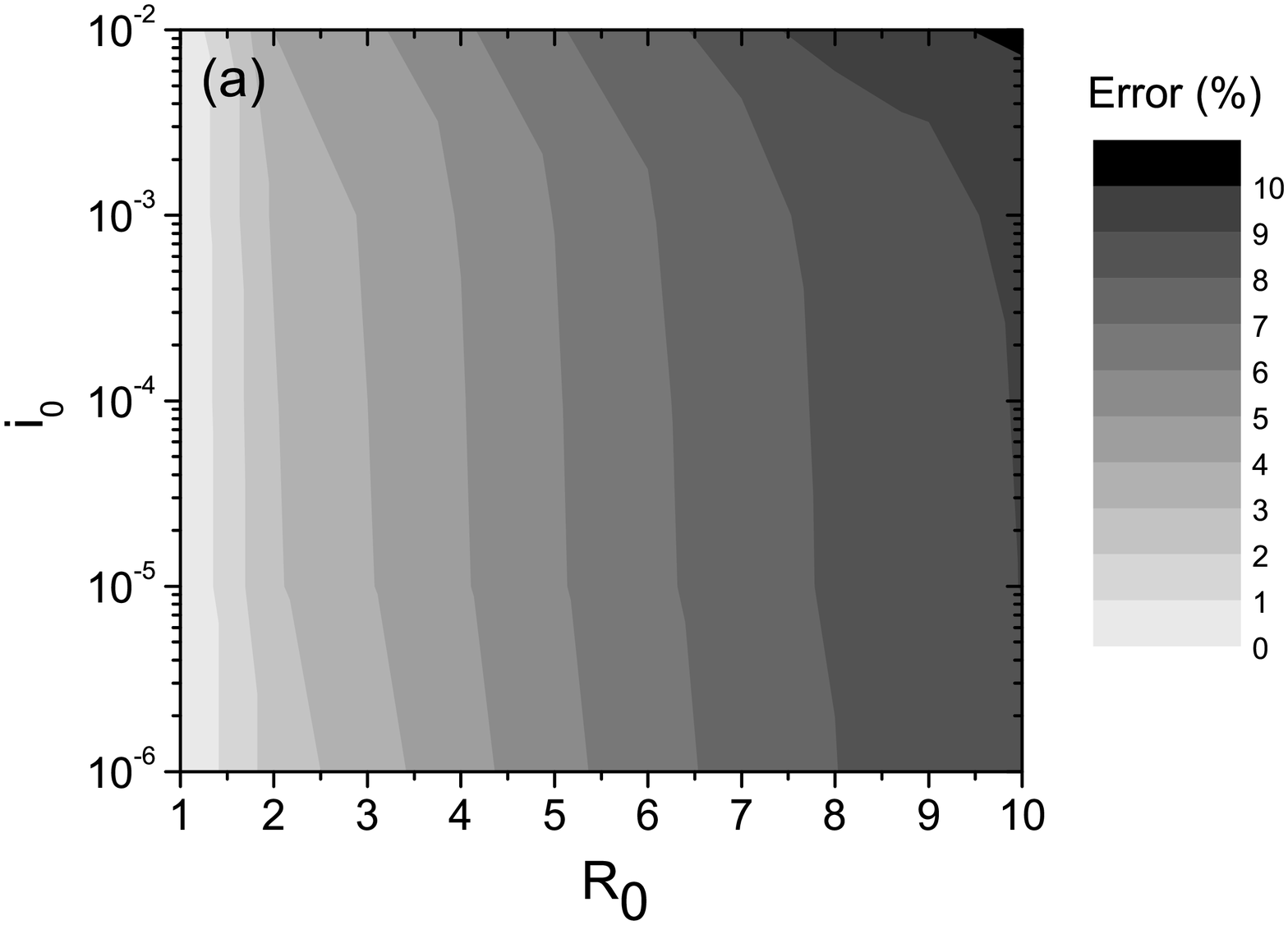}}}
\centerline{\subfigure{\includegraphics[width=8.5cm]{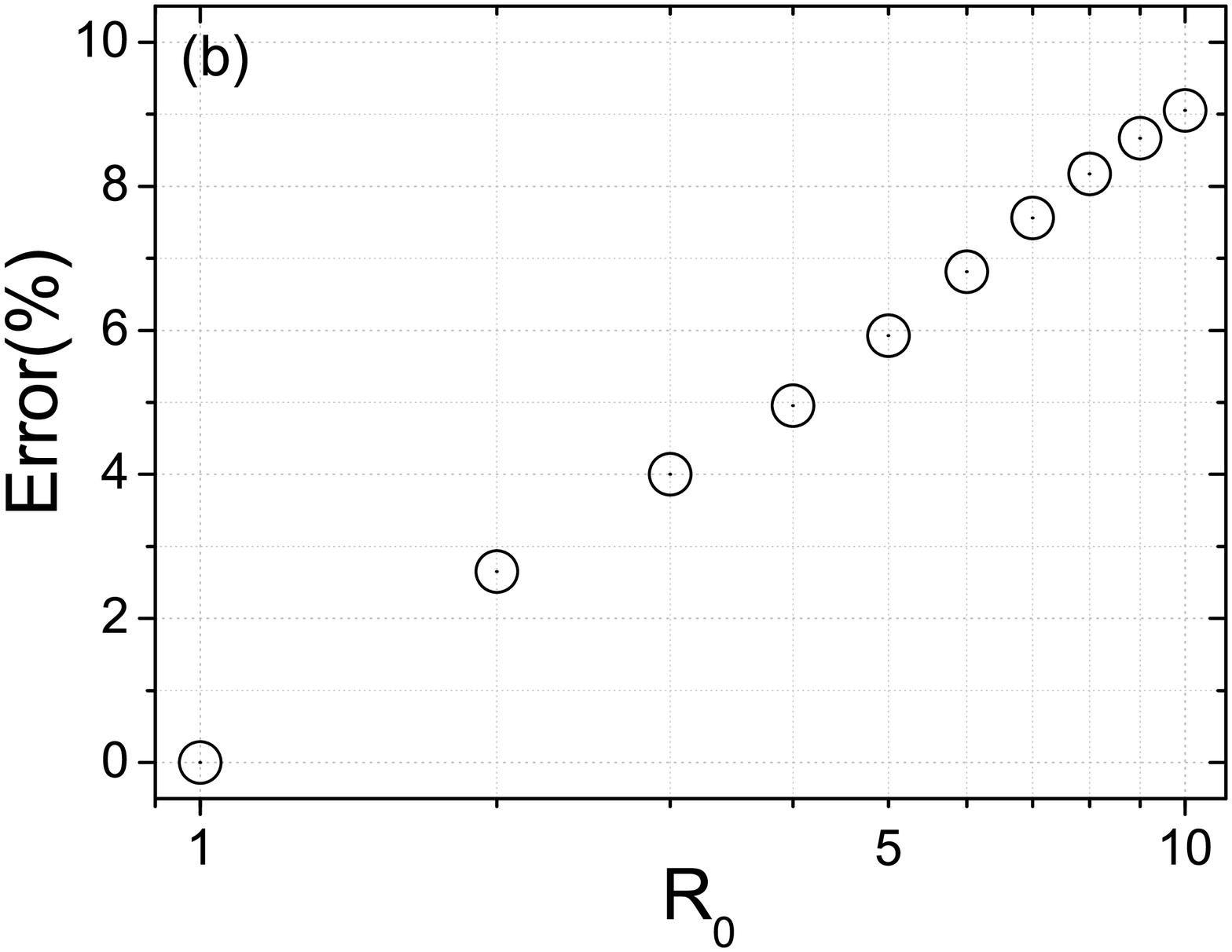}}}
\centerline{\subfigure{\includegraphics[width=8.5cm]{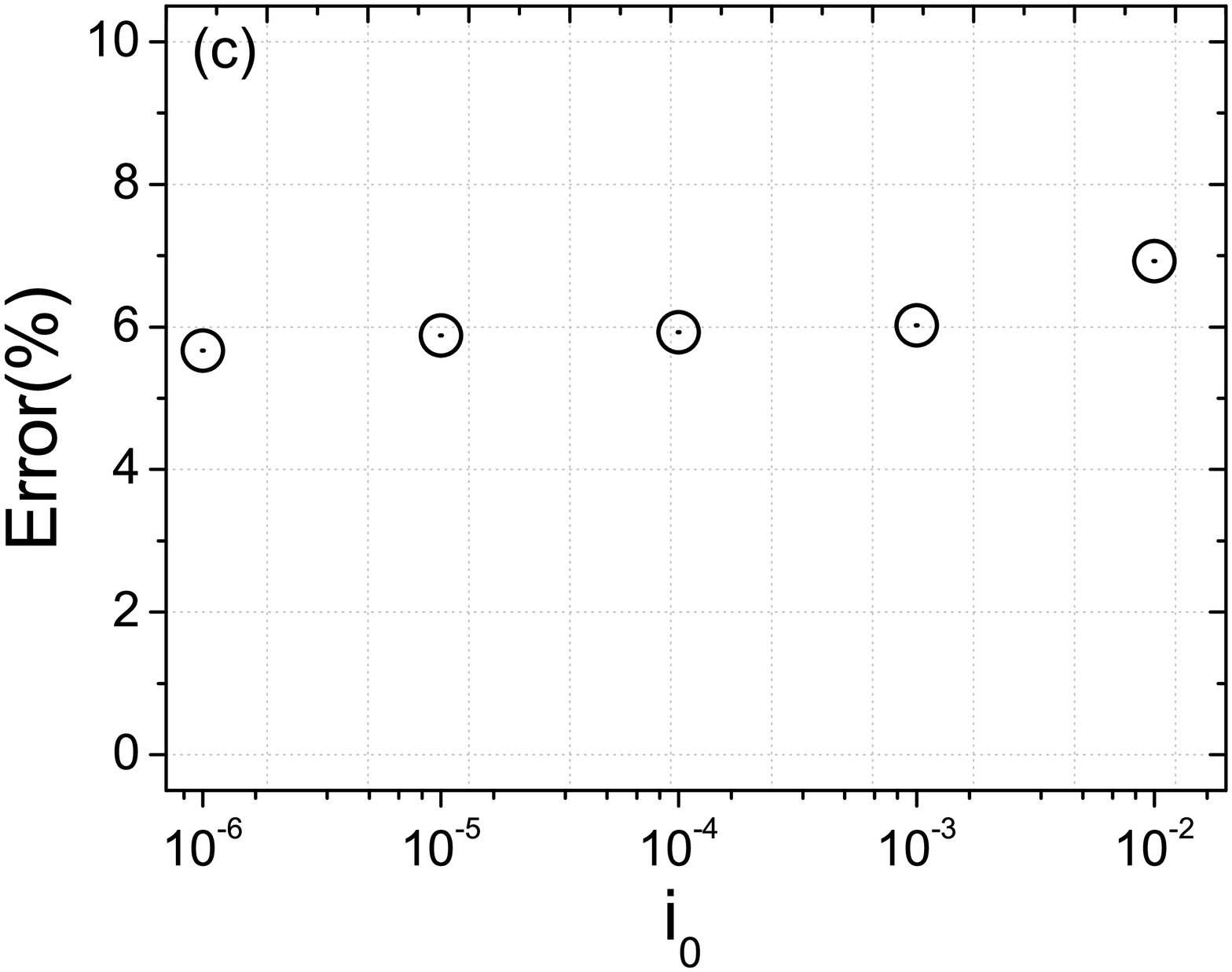}}}
\caption{(a) Percentage relative error on removed $R(t)$ as function
  of $R_0$ and $i_0$. Two cross-section cuts or views of the previous
  curve: (b) cut at $i_0=10^{-3}$ and (c) cut at $R_0 = 5$.}
\label{fig:ErroSIR}
\end{figure}
This way, based on Eq.~(\ref{eq:lw}) and~(\ref{eq:tauf}), the SIR
dynamics can be algebraically obtained given the initial conditions:
$I(0)=i_{0}$, $S(0)=1-i_{0}$, and the basic reproductive number $R_0$.
Note that the function $\omega(\tau)$ has no inverse (which would be
the ideal situation), however that does not keep us of getting an
explicit representation of the dynamics, i.e to obtain explicit
functions for the three quantities $S$, $I$, and
$R$. Fig.~\ref{fig:sir} shows a comparison between the numerical
integration of the differential equations (Eq.~(\ref{eq:sir})) and the
(numerically computed) proposed algebraic solution (Eq.~(\ref{eq:lw})
and~(\ref{eq:tauf})).

\section{Inspection of the Algebraic Solution}
More detailed inspection of that figure as it displayed in
Fig.~\ref{fig:sir}(b) indicates that the agreement is not perfect as a
consequence of the approximation done in Eq.~(\ref{eq:lw}).  In order
to quantify the error of the algebraic solution, we compute the
relative differences between removed $R(t)$ obtained from the
numerical solution of the differential equations (Eq.~\ref{eq:sir})
and from the present solution (Eq.~(\ref{eq:lw}) and~(\ref{eq:tauf})),
averaged up to the asymptotic state, as follows:
\begin{equation}
Error =
\frac{1}{N}\sum_{k=1}^{N}\frac{\left|R_N(\tau_k)-R_A(\tau_k)\right|}{R_N(\tau_k)}\times
100
\end{equation}
where $N=int(\tau_F/d\tau)$, $int(x)=\max\{m \in Z | m \leq x \} $,
and $\tau_k=(k-1)d\tau$.  In order to evaluate such error we take
$d\tau=10^{-3}$ (time step of the numerical integration) and
$R(\tau_F)=0.99R(\infty)$ (asymptotic or final time). The labels $N$
and $A$ refer to numerical and algebraic solutions, respectively.  In
Fig.~\ref{fig:ErroSIR}(a) we present the percentage error as defined
above versus the basic reproductive number $R_0$ and initial fraction
of infected subjects $i_0$. It can be seen that the error is sensitive
to both $R_0$ and the initial fraction of infected $i_0$. In
particular, if $i_0 < 10^{-3}$, the $Error$ is proportional to
$\log{R_0}$ (see Fig.~\ref{fig:ErroSIR}(b)). Furthermore, the $Error$ is
practically constant in relation to $i_0$ (see
Fig.~\ref{fig:ErroSIR}(c)). Using linear regression method we estimate
that $Error(R_0,i_0) \simeq 9\log{R_0}$.

\section{Consequences of the Algebraic Solution}
\begin{figure}[!t]
\centerline{\subfigure{\includegraphics[width=8.5cm]{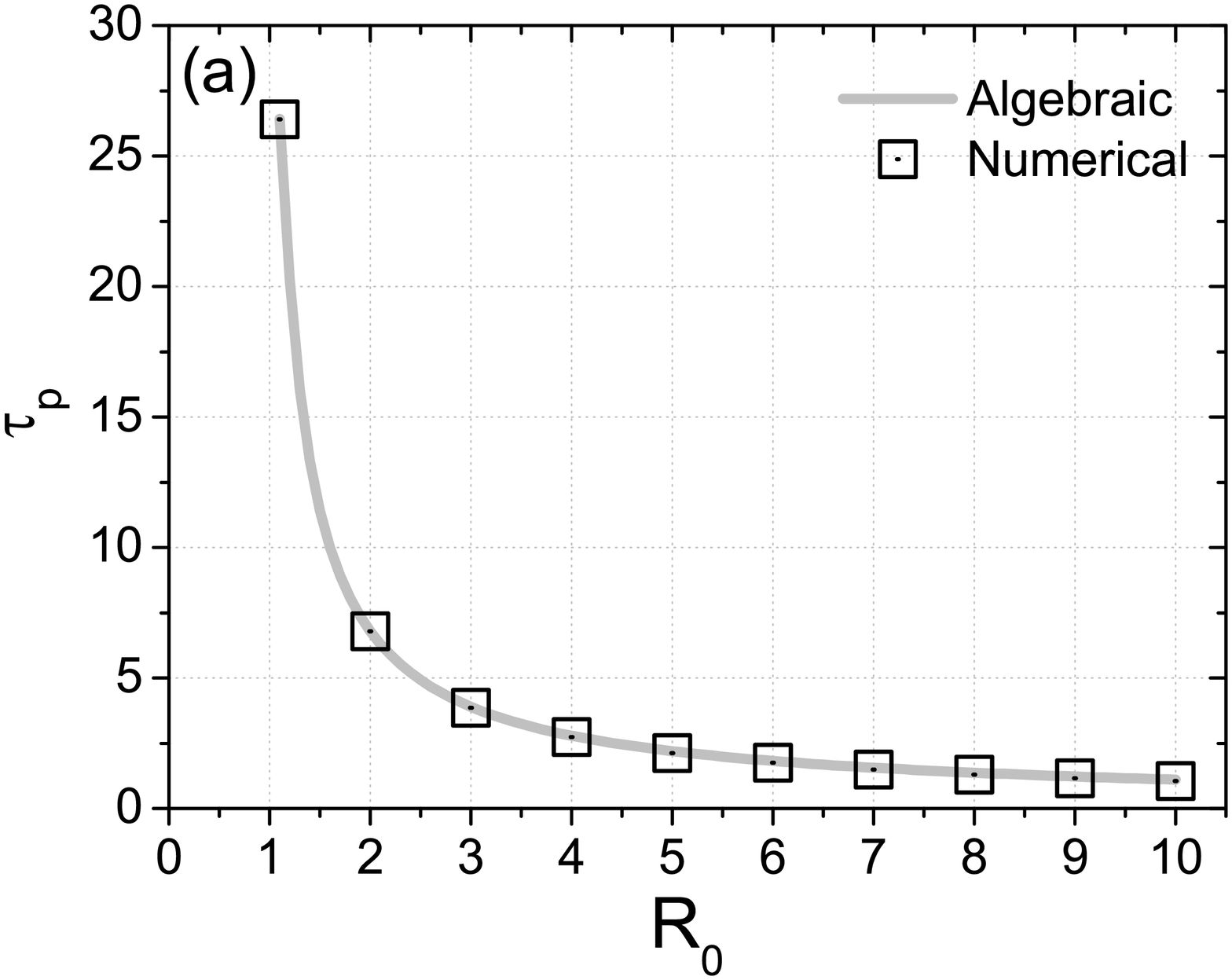}}}
\centerline{\subfigure{\includegraphics[width=8.5cm]{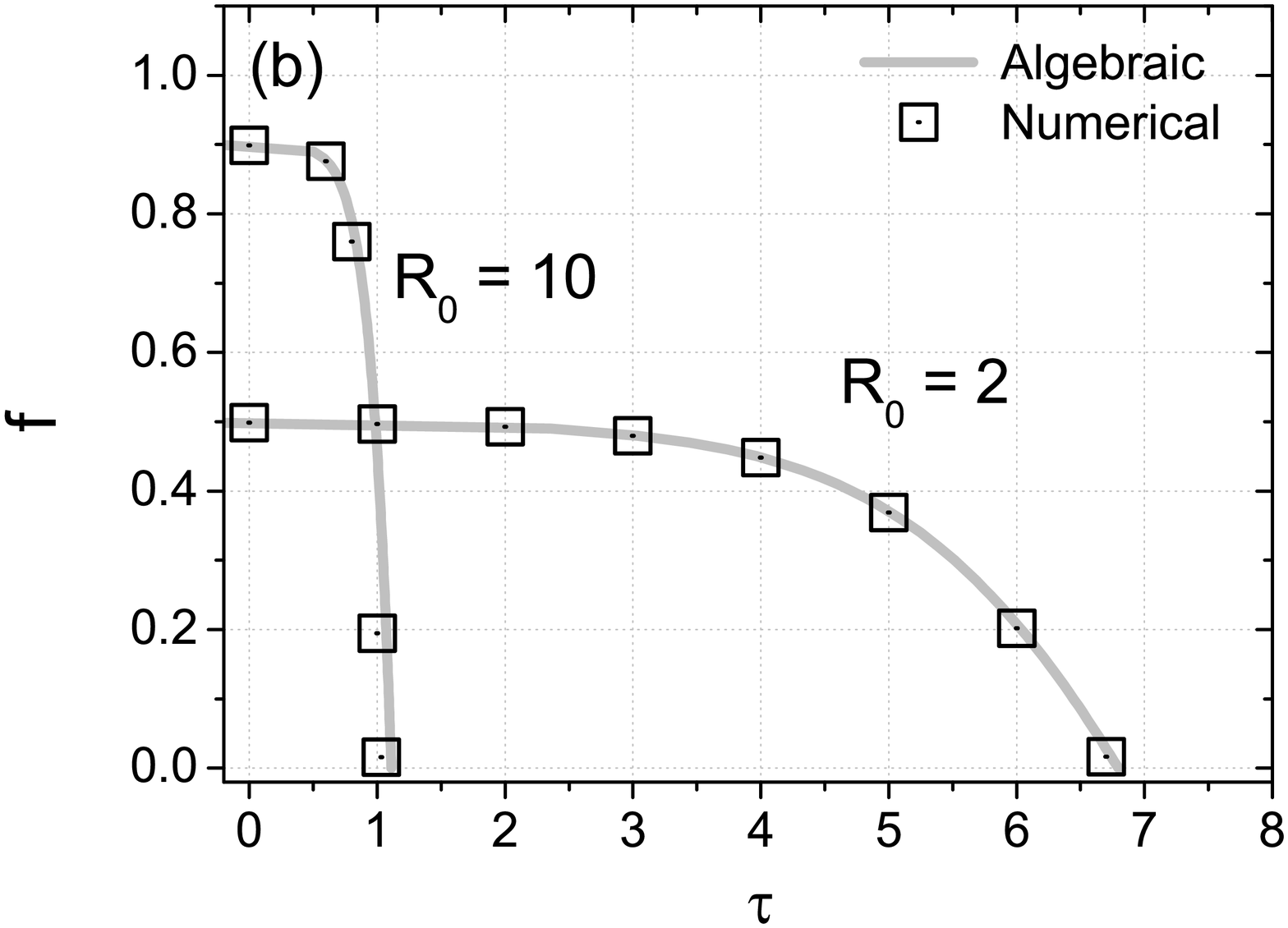}}}
\caption{Comparison between algebraic and numerical solutions for (a),
  time of the infection peak, and (b), fraction of immunized. Initial
  condition $i_0=10^{-3}$.}
\label{fig:Tp}
\end{figure}

From the previous results, we can arrive at an analytic expression for
the time of the infection peak, $\tau_p$.  For that purpose we plug
the expressions $w(\tau_p)=R_0-\ln \left[R_0 S(0)\right]$ into the
equation Eq.~(\ref{eq:tauf}), we obtain
\begin{eqnarray}
\tau_p \simeq&&
\frac{\ln{\left[1-\frac{\ln\left[R_0S(0)\right]}{R_0
        (1-S(\infty))}\right]}}{R_0S(0) S(\infty)-1} \nonumber
\\ &+&\frac{\ln{\left[1+\frac{\ln[R_0S(0)](R_0S(0)-1)}{R_0(1-S(0))}\right]}}{{R_0S(0)\exp{\left[\frac{R_0(1-S(0))}{R_0S(0)-1}\right]}-1}}
\label{eq:Tp}
\end{eqnarray}
In the Fig.~\ref{fig:Tp}(a) we present the $\tau_p$ as function of
$R_0$. Note that previous equation is according with the numerical
solution (numerical integration Eq.~(\ref{eq:sir})). This last result,
together with the expression for the basic reproductive number,
$R_{0}$, the asymptotic fraction of removed subjects $R(\infty)$, and
the maximum fraction of infected people, $I_{max}$, i.e,
\begin{eqnarray}
R_0&=& \frac{\beta}{\gamma} \\ I_{max} &=&
1-\frac{1}{R_0}\left(1+\ln[R_0 S(0)]\right)\\ R(\infty)&=&
1-S(0)e^{-R_0 R(\infty)},
\end{eqnarray}
represent in algebraic form the principal quantities of the SIR model
(presented in the original article of Kermack-McKendrick).

It is worth noting that the above results allow us to obtain close
expression for other important quantities related to extensions of the
Kermack-McKendrick formulation, such is the case of the immunization
necessary to stop an epidemic.  According to Anderson and
May~\cite{Anderson}, the fraction of the population $f$ that have to
be vaccinated in order to stop the spread of the disease in the
population, is $f=1-1/(R_0 S(0))$. However, in many cases, we can not
immunize the population before the start of the infection
process. Thus, we can generalize the expression of the fraction of
immunized with a time dependence, $f(\tau)=1-1/(R_0 S(\tau))$. Note
that with the assistance of the Eq.~(\ref{eq:lw}) and
Eq.~(\ref{eq:tauf}), we have algebraic expression for the fraction of
individuals that should be immunized in time $\tau$ for the disease to
be extinct. To do so, just make the substitution
$w(\tau)=R_0-\ln\left[R_0 S(0)\left(1-f(\tau)\right)\right]$
in the Eq.~(\ref{eq:tauf}), results
\begin{eqnarray}
 \tau\simeq&&
 \frac{\ln{\left[1-\frac{\ln\left[R_0 S(0)\left(1-f(\tau)\right)\right]}{R_0
        (1-S(\infty))}\right]}}{R_0 S(0) S(\infty)-1} \nonumber
\\ &+&\frac{\ln{\left[1+\frac{\ln[R_0 S(0)\left(1-f(\tau)\right)]
(R_0 S(0)-1)}{R_0(1-S(0))}\right]}}{{R_0S(0)\exp{\left[\frac{R_0(1-S(0))}{R_0S(0)-1}\right]}-1}}
         \label{eq:FI}
\end{eqnarray}
Fig.~\ref{fig:Tp}(b) shows the dynamics of the fraction of
immunized. Note that in the initial moments of spread of the epidemic
$f(\tau)$ remains constant. However, as time progresses, the fraction
of the population to be immunized falls significantly as it approaches
the peak time of infection. Evidently, $f(\tau) = 0$ if $\tau \geq
\tau_p$ since that $S(\tau)\leq 1/R_0$.

\section{Final Remarks}
We have presented an algebraic solution for an important and long
standing problem of the mathematical biology, which is the solution of
the differential equations of the SIR model, as first presented by
Kermack and McKendrick.  In this solution, the dynamics of the
fraction of susceptible, infected, and removed are given explicitly in
terms of a time-dependent expression $w(\tau)$ (Eq.~(\ref{eq:lw})).
Using well justified approximations, we finally arrive to a
transcendental expression for $w(\tau)$ (Eq.~(\ref{eq:tauf})), in
terms of which all the dynamic quantities can be expressed.  We
verified that the present general solution is in excellent agreement
with numerical solutions of the same equations over the entire dynamic
for different threshold ($R_0$) values. We showed that the difference between
these solutions (algebraic and numeric) is proportional to the
logarithm of the epidemic threshold $R_0$. In this sense, the error in
the proposed solution is small and grows slowly as the value of $R_0$.
Lastly, this study enables us to represent analytic expressions for
the time of the infection peak (Eq.~(\ref{eq:Tp})) and
fraction of immunized (Eq.~(\ref{eq:FI})).

\section{Acknowledgments}
We acknowledge support from the Brazilian agencies CNPq and CAPES and
partial support from CNPq project \#551974/2011-7.

\bibliographystyle{phcpc}

\end{document}